 \documentclass[final,5p,times,twocolumn, authoryear]{elsarticle}

\usepackage{amssymb}
\usepackage{lipsum}
\usepackage{amsmath}
\usepackage{subcaption}
\usepackage{hyperref}

\usepackage{lineno}

\newcommand{\hess}{H.\,E.\,S.\,S.\,}

\journal{High Energy Astrophysics}

\begin{document}

\begin{frontmatter}



\title{Multiwavelength study of non-thermal emission in the Swift J1834.9–0846/W41 region}


\author[01]{Manoel F. Sousa}
\affiliation[01]{organization={Instituto de Física de São Carlos, Universidade de São Paulo},
            addressline={Av. Trabalhador São-carlense 400}, 
            city={São Carlos},
            postcode={13566-590}, 
            state={SP},
            country={Brazil}}

\author[02,03,04,05,06]{Rita C. Dos Anjos}
\affiliation[02]{organization={Departamento de Engenharias e Exatas, Universidade Federal do Paraná (UFPR)},
            addressline={Pioneiro, 2153}, 
            city={Palotina},
            postcode={85950-000}, 
            state={PR},
            country={Brazil}}

\affiliation[03]{organization={Programa de Pós-Graduação em Física e Astronomia, Universidade Tecnológica Federal do Paraná},
            city={Curitiba},
            postcode={80230-901}, 
            state={PR},
            country={Brazil}}

\affiliation[04]{organization={N\'ucleo de Astrof\'{\i}sica e Cosmologia (Cosmo-Ufes), Universidade Federal do Esp\'irito Santo},
            city={Vit\'oria},
            postcode={29075-910}, 
            state={ES},
            country={Brazil}}

\affiliation[05]{organization={Programa de pós-graduação em Física, Universidade Estadual de Londrina (UEL)},
            addressline={Rodovia Celso Garcia Cid Km 380}, 
            city={Londrina},
            postcode={86057-970}, 
            state={PR},
            country={Brazil}}

\affiliation[06]{organization={Programa de Pós-Graduação em Física Aplicada, Universidade Federal da Integração Latino-Americana},
            city={Foz do Igua\c{c}u},
            postcode={85867-670}, 
            state={PR},
            country={Brazil}}

\author[01]{Vitor de Souza}

\begin{abstract}

We investigate the origin of non-thermal emission from the Swift~J1834--0846/W41 region by modeling its broadband spectral energy distribution from radio to TeV energies within leptonic and lepto-hadronic frameworks using Markov Chain Monte Carlo sampling. Motivated by morphological studies of HESS~J1834--087 suggesting a two-component TeV structure, we explore a single extended source scenario and a configuration comprising a central point-like component embedded within extended emission. Purely leptonic models are disfavored in both scenarios by unrealistically low magnetic field strengths, whereas lepto-hadronic solutions yield field intensities and non-thermal energy budgets consistent with an evolved supernova remnant undergoing efficient cosmic-ray acceleration. In the two-component scenario, hadronic interactions dominate the extended TeV emission from W41, while the central excess is well described by a leptonic magnetar wind nebula powered by Swift~J1834--0846, implying a short initial spin period of $P_0 \lesssim 0.2$~s. Simulated observations with the Cherenkov Telescope Array Observatory show that 30~h exposures will discriminate between the two morphological configurations and extend spectral measurements beyond $\sim$10~TeV.

\end{abstract}



\begin{keyword}
gamma rays \sep magnetar \sep supernova remnant \sep pulsar wind nebula \sep acceleration of particles \sep cosmic rays



\end{keyword}

\end{frontmatter}




\section{Introduction}   \label{introduction}

The region encompassing Swift~J1834--0846 and the supernova remnant (SNR) W41 provides an important laboratory for investigating the interplay among compact objects, their surrounding nebulae, and evolved remnants at high energies. Situated in a complex and gas-rich environment in the Galactic plane~\citep{2007ApJ...657L..25T, 2008AJ....135..167L}, 
W41 has long been regarded as a potential site of efficient particle acceleration~\citep[see, e.g.,][]{2013ApJ...773L..19F,2015A&A...574A..27H}. In particular, the detection of OH maser emission at 1720~MHz provides compelling evidence for a physical interaction between the SNR shock and the adjacent giant molecular cloud G23.0--0.4~\citep{2013ApJ...773L..19F}, suggesting that proton--proton collisions may be especially 
efficient in this environment. The presence of Swift~J1834--0846, one of the few magnetars spatially associated with TeV emission, adds an additional level of physical complexity to the system. Moreover, its positional coincidence with the TeV gamma-ray source HESS~J1834--087~\citep{2018A&A...612A...2H} further highlights the importance of this 
region for understanding the origin of very-high-energy (VHE) gamma-ray radiation. Notably, detailed morphological studies of HESS~J1834--087 have revealed hints of a two-component TeV structure, comprising a central point-like source superimposed on extended emission~\citep{2015A&A...574A..27H}, which naturally motivates a multi-component interpretation of the observed gamma-ray signal.

Magnetars, characterized by ultra-strong magnetic fields reaching up to $10^{15}$~G and transient high-energy activity, are not typically regarded as efficient accelerators of particles up to TeV energies \citep{2017ApJ...835...30L}. Nevertheless, growing observational 
evidence suggests that at least some magnetars may power extended nebulae capable of accelerating leptons to relativistic energies 
\citep{2016ApJ...824..138Y, 2016RAA....16..143T, 2017MNRAS.464.4895G}. In this context, Swift~J1834--0846 is of particular interest because it is associated with diffuse X-ray emission interpreted as a magnetar wind nebula \citep[MWN;][]{2016ApJ...824..138Y}, broadly analogous to classical pulsar wind nebulae (PWNe). If such systems are capable of producing TeV emission, they could provide valuable insights into the energy budget and spin evolution of young magnetars.

At the same time, SNRs remain among the most promising candidates for the acceleration of Galactic cosmic rays up to at least the knee of the cosmic-ray spectrum~\citep[see, e.g.,][]{1988ApJ...333L..65V, 2004A&A...424..747P, 2012JCAP...07..038C,2013MNRAS.431..415B, 2021MNRAS.504.6096M, 2023MNRAS.519..136V, 2024PhRvD.110d3046K, 2025MNRAS.luana}. In middle-aged remnants interacting with dense molecular material, proton--proton collisions can efficiently produce neutral pions that subsequently decay into gamma rays, leading to extended TeV emission spatially correlated with the remnant shell or nearby molecular clouds. The detection of extended VHE gamma-ray emission in the vicinity of W41 therefore raises the possibility that a substantial fraction of the observed radiation originates from hadronic interactions 
within the SNR environment \citep[see, e.g.,][]{2013ApJ...773L..19F,2015A&A...574A..27H}.

The coexistence of a magnetar, a candidate MWN, and an evolved SNR within the same region naturally motivates a multi-component interpretation of the gamma-ray emission. From a theoretical perspective, this configuration provides an opportunity to investigate multiple particle-acceleration channels operating simultaneously. In particular, determining whether 
the TeV radiation is dominated by leptonic and/or hadronic processes in the SNR, or by leptonic emission from a central nebula powered by the magnetar, bears directly on the possible contribution of magnetars to the high-energy particle population in the Galaxy.

Motivated by these considerations, we present a comprehensive theoretical investigation of particle acceleration and non-thermal emission in the Swift~J1834--0846/W41 region~\citep[][]{2011GCN.12253....1D,2012ApJ...748...26K,2017MNRAS.464.4895G}. We perform 
detailed multiwavelength spectral modeling covering the radio to TeV energy range, exploring both purely leptonic and lepto-hadronic scenarios. The broadband spectral energy distribution (SED) is modeled using Markov Chain Monte Carlo (MCMC) sampling within the \texttt{Naima}~framework \citep{naima}, allowing us to derive posterior distributions for the underlying particle populations and their radiative components, including synchrotron radiation, inverse-Compton (IC) scattering, and gamma-ray emission from neutral-pion decay. The modeling follows standard formalisms for non-thermal radiation processes~\citep[e.g.,][]{1979rpa..book.....R,1970PhRvD...1.1596B,2006PhRvD..74c4018K}, enabling us 
to quantify the relative contribution of each emission channel and to assess whether the observed gamma-ray emission is predominantly leptonic or hadronic~\citep{Landshoff:1971xb,2019scta.book.....C}.

In addition, we investigate the observational prospects for the Cherenkov Telescope Array Observatory (CTAO) in the Swift~J1834--0846 region. To this end, we performed dedicated three-dimensional simulations of gamma-ray maps and derived the corresponding flux points using the \texttt{Gammapy} analysis framework~\citep{2023A&A...678A.157D, acero_2025_17814297}. These simulations allow us to assess the capability of CTAO to resolve the morphology of the TeV emission.

The results presented here have broad implications for understanding the role of magnetars and their nebulae as potential high-energy particle accelerators. By characterizing the dominant radiation mechanisms and comparing them with those expected from SNR-driven acceleration, our 
study contributes to a more comprehensive picture of non-thermal processes in complex compact-object environments. Moreover, the predictions obtained from our modeling provide a useful theoretical framework for interpreting future high-sensitivity gamma-ray observations with CTAO.

This paper is organized as follows. In Section~\ref{sec:swift_reg}, we introduce the main observational properties and previous constraints for the Swift~J1834--0846/W41 region. Section~\ref{sec:modeling} describes the multiwavelength spectral modeling performed under both leptonic and lepto-hadronic frameworks, considering two morphological scenarios for the 
TeV emission, namely a single extended component and a two-component configuration. The results of these analyses are presented in Section~\ref{sec:results}, where the extended emission is interpreted as arising from the SNR shock, while the central point-like component 
is examined in the context of a MWN; the physical implications of these scenarios are also discussed. In Section~\ref{sec:CTAO}, we investigate the prospects for testing these interpretations with future observations by the CTAO. Finally, Section~\ref{conc} summarizes our main conclusions.

\section{Swift~J1834-0846/W41 region}      \label{sec:swift_reg}

\begin{figure}
\centering
\includegraphics[width=0.49\textwidth]{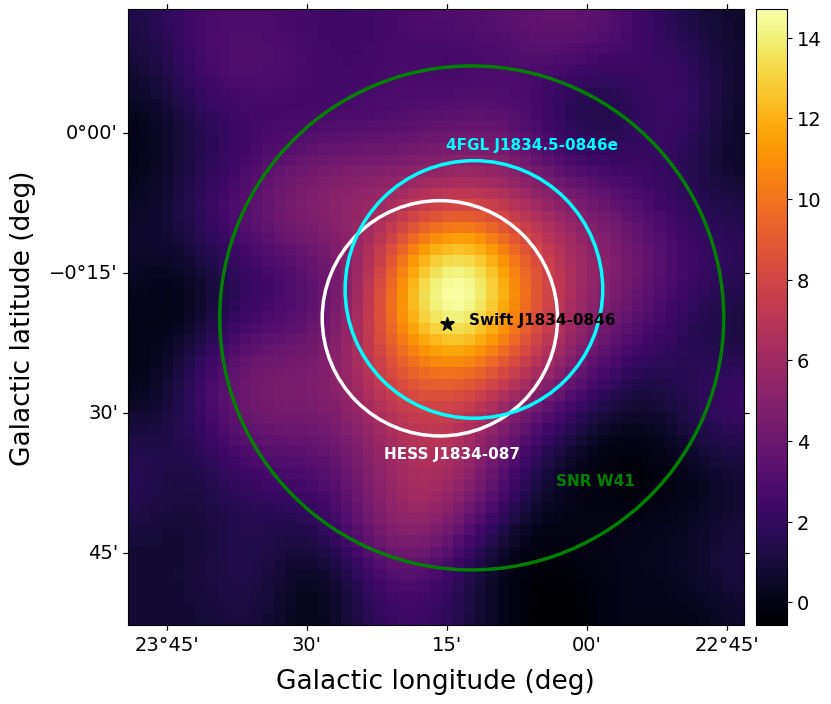}
\caption{H.E.S.S. VHE gamma-ray significance map of the region surrounding 
SNR~W41 and Swift~J1834--0846~\citep{2018A&A...612A...1H}, smoothed with a 
Gaussian kernel of width $0.05^{\circ}$. The black star marks the position of the magnetar Swift~J1834--0846~\citep{2014ApJS..212....6O}. The green circle indicates the radio extent of W41, with a radius of $27'$ ~\citep{2007ApJ...657L..25T, 2025JApA...46...14G}. The cyan circle represents the GeV counterpart detected by \textit{Fermi}-LAT~\citep{2022ApJS..260...53A}, while the white circle corresponds to the $1\sigma$ extension of the TeV source as measured by H.E.S.S.~\citep{2018A&A...612A...1H}. The map illustrates the spatial coincidence between the extended TeV emission of HESS~J1834--087 and the radio boundary of the SNR, motivating the multi-component interpretation 
explored in this work.}
\label{fig:map_region}
\end{figure}

Swift~J1834--0846 is a magnetar situated approximately 4~kpc 
away in the constellation Scutum, at the center of the supernova remnant W41 (see Fig.~\ref{fig:map_region}). Following its discovery on August 7, 2011, by NASA's Swift satellite during an X-ray outburst, extensive X-ray and gamma-ray studies have been conducted to understand its emission properties and surrounding environment. Its observed spin period is $2.48$~s, with a period derivative of $\dot{P} = 7.96 \times 10^{-12}~\text{s\,s}^{-1}$ indicating a surface magnetic field strength of $10^{14}$~G \citep[see][]{2011GCN.12253....1D, 2012ApJ...748...26K, 2016ApJ...824..138Y, 2017MNRAS.464.4895G}.

Although the positional coincidence of Swift~J1834–0846 with the projected center of W41 strongly suggests a physical association, this link remains debated. One of the main challenges lies in reconciling the different age estimates: the magnetar’s characteristic spin-down age is only $\sim 4.9$~kyr \citep{2014ApJS..212....6O}, whereas the remnant’s estimated age lies in the range $60$–$200$~kyr \citep{2007ApJ...657L..25T, 2015A&A...574A..27H}. Reconciling this discrepancy requires either a reassessment of the remnant's dynamical age or the assumption that the magnetar's characteristic age does not reliably reflect its true evolutionary timescale, which often occurs due to the complex torque history of magnetars \citep{1993ApJ...408..194T, 2020MNRAS.494.4838L}.

A further difficulty concerns the magnetar’s proper motion. Its current position close to the geometric center of W41 implies that its motion must be directed mostly along the line of sight. Otherwise, for a typical pulsar velocity of $\sim 400$~km s$^{-1}$ \citep{2005MNRAS.360..974H}, Swift~J1834–0846 would have already traversed toward the edge of the remnant over the past $60$~kyr, in conflict with observations. High-precision astrometric measurements of its motion will therefore be essential to determine its trajectory and to test whether Swift~J1834–0846 and W41 indeed share a common origin in the same supernova explosion.

Observations conducted with the European Space Agency’s \textit{XMM-Newton} X-ray observatory have provided significant insights into Swift~J1834–0846. Two deep \textit{XMM-Newton} observations, performed in 2014, confirmed the presence of extended X-ray emission centered around the magnetar. The emission exhibits an elongated morphology stretching southwestward from the magnetar position. Spectral analysis of extended emission revealed that it is best described by a highly absorbed power-law model with a photon index of \(\Gamma_\gamma = 2.2 \pm 0.2\) and a hydrogen column density of $N_H = 8.0 \times 10^{22}$~cm${-2}$. The flux remained stable across multiple observations spanning nine years, measured at $F_{0.5-10 keV} = 1.3 \times 10^{-12}$~erg~cm$^{-2}$~s$^{-1}$, suggesting a persistent and non-variable source \citep[see][]{2016ApJ...824..138Y, 2017MNRAS.464.4895G}.

This extended X-ray structure is now interpreted as a MWN, making Swift~J1834–0846 the first known magnetar to exhibit such a feature. The properties of the nebula closely resemble those of the PWNe, but with an exceptionally high radiative efficiency, estimated at \(\eta_X = L_X / \dot{E} \approx 0.1\). This efficiency level is significantly higher than typical PWNe, suggesting an additional energy component -- possibly linked to magnetar outflows and burst activity -- contributing to its luminosity \citep{2016ApJ...824..138Y}. However, \cite{2017ApJ...835...54T} has proposed an alternative explanation, arguing that the nebula’s size and emission characteristics indicate it could be rotationally powered. In this scenario, compression caused by environmental interactions leads to adiabatic heating, enhancing its brightness. The magnetic field strength derived within the nebula is estimated to range between $11 \, \mu$G and $30 \, \mu$G \citep{2016ApJ...824..138Y}.

In the gamma-ray regime, the region contains the TeV source HESS J1834-087 (see Fig.~\ref{fig:map_region}), which is probably produced by interactions between cosmic rays from the supernova remnant W41 and nearby molecular clouds, rather than direct emission from the magnetar. The absence of a significant non-thermal X-ray counterpart to the TeV emission strengthens the case for a hadronic origin, although a possible contribution from a PWN component cannot be completely ruled out \citep[see e.g.,][]{2016ApJ...824..138Y}. \cite{2017MNRAS.464.4895G} argue that spectral analysis reinforces this interpretation, suggesting that the GeV and TeV emission components are best explained by interactions between cosmic rays accelerated at the SNR shock and dense molecular clouds in the region.

\section{Source modeling}   \label{sec:modeling}

We investigate the region around the magnetar Swift~J1834-0846 applying MCMC techniques to fit radiative models via the \texttt{Naima} software \citep{naima}. We investigate whether the observed high- and very-high-energy (VHE) gamma-ray emission can be explained by leptonic or lepto-hadronic scenarios. In these models, we consider particle acceleration in SNR~W41 via the diffusive shock acceleration mechanism, as well as the acceleration of relativistic electrons at the wind termination shock beyond the light cylinder. In the latter case, the accelerated particles originate from the MWN directly powered by Swift~J1834–0846.

The particle energy distributions were modeled assuming a distance of $4.2$~kpc to the Swift~J1834–0846 region \citep{2016ApJ...824..138Y}. This estimate is supported by the close spatial association of Swift J1834–0846 with the geometric center of SNR~W41, whose distance was estimated to be $4.0 \pm 0.2$~kpc by \citet{2008AJ....135..167L}. In the context of synchrotron emission, where electrons experience significant radiative energy losses on timescales $\gtrsim 1$~kyr, the underlying particle energy distribution is often expected to follow an exponential cutoff broken power law (\textit{ECBPL}), which reflects the cooling break in the particle spectrum. In contrast, protons, owing to their much larger mass, do not experience significant synchrotron or inverse Compton (IC) losses. Although they may lose energy through inelastic proton–proton interactions leading to pion production, the corresponding cooling timescale is considerably longer than the estimated age of the source. For this reason, the proton spectrum is typically modeled with a simpler exponential cutoff power-law (\textit{ECPL}) distribution. Accordingly, we adopted an \textit{ECBPL} model to represent the electron population in both leptonic and lepto-hadronic scenarios, and an \textit{ECPL} model to describe the proton population in the lepto-hadronic framework. The {\it ECPL} model is given by:

\begin{equation}
\Phi(E) = \Phi_{0} \left( \frac{E}{E_{0}} \right)^{-\Gamma} \exp\left( -\frac{E}{E_{\mathrm{cut}}} \right),
\label{eq:ecpl}
\end{equation}
while the {\it ECBPL} model is expressed as:
\begin{equation}
    \Phi (E) = \Phi_{0} 
    \begin{cases}
    \left(\dfrac{E}{E_b}\right)^{-\Gamma_1}, & E < E_b, \\[1.2em]
    \left(\dfrac{E}{E_b}\right)^{-\Gamma_2} \,
    \exp\!\left(-\dfrac{E}{E_{\rm cut}}\right), & E \geq E_b,
    \end{cases}
\label{eq:ecbpl}
\end{equation}
where \( \Phi (E) \) denotes the flux at energy $E$, \( \Phi_{0} \) is the flux normalization at the reference energy \( E_{0} \), \( \Gamma \) is the spectral index, \( E_{\mathrm{cut}} \) is the gamma-ray energy cutoff, $E_b$ is the break energy, and $\Gamma_1$ and $\Gamma_2$ are the spectral indices before and after the break, respectively. In addition, within the lepto-hadronic framework, we investigated the ratio between the normalization constants of the electron and proton energy distributions, defined as $K_{\rm ep} = \Phi_{\rm 0,e}/\Phi_{\rm 0,p}$ \citep{2012A&A...538A..81M}. The normalization is evaluated at a reference energy of $1$~TeV. This procedure allows us to constrain the relative proton contribution required to account for the observed gamma-ray emission.

The radiation field adopted in the models includes contributions from the CMB, a far-infrared (FIR) component with a temperature of $T = 25$~K and an energy density of $u = 0.5$~eV cm$^{-3}$, and a near-infrared (NIR) component with $T = 3000$~K and $u = 1.0$~eV cm$^{-3}$ \citep{2017ApJ...835...54T}. For the ambient gas density, \citet{2007ApJ...657L..25T} derived an average interstellar density of $\sim 6$ cm$^{-3}$ from HI column-density maps associated with SNR W41. We adopt this value in our analysis, while noting that substantial local density variations are suggested by $^{12}$CO observations tracing a giant molecular cloud that may be interacting with the remnant \citep{2007ApJ...657L..25T,2015ApJ...811..134S}. We assume that the radio-to-X-ray emission arises from synchrotron radiation produced by relativistic electrons interacting with the magnetic field, while VHE gamma-ray emission results from IC scattering of background radiation and from neutral pion decay.

In order to model the gamma-ray spectrum, we used observational data from the \textit{Fermi}-LAT and \hess instruments \citep{2018A&A...612A...1H, 2015A&A...574A..27H, 2022ApJS..260...53A}. A detailed analysis of the \hess observations of the Swift~J1834–0846 region revealed a two-component morphology for HESS~J1834–087: a central point-like source and an extended component \citep{2015A&A...574A..27H}. Motivated by this, we considered two scenarios to investigate the relative contributions of the SNR and MWN to the VHE emission.

In the first scenario, the entire TeV emission is modeled as a single source in order to assess whether the SNR alone can account for the observed spectrum, without invoking an additional emission component. In the second scenario, the analysis distinguishes between the extended TeV emission and the central point-like TeV component, attributing the latter to the MWN, while the SNR is assumed to dominate the extended TeV emission.

In the SNR scenario, the synchrotron component is constrained by radio measurements and by an upper limit derived from X-ray observations. The radio data are taken from the Very Large Array (VLA) Galactic Plane Survey (VGPS) at $330$~MHz and $1420$~MHz \citep{1992AJ....103..943K, 2007ApJ...657L..25T}, while the X-ray upper limit is obtained from {\it XMM-Newton} observations associated with the MWN \citep{2016ApJ...824..138Y}. In the MWN scenario, the synchrotron emission is constrained by X-ray observations of the MWN, which exhibit a photon index of $\Gamma_{\rm ph} = 2.2 \pm 0.2$ \citep{2016ApJ...824..138Y}, whereas the radio measurements of the SNR are treated as upper limits.

\begin{table*}
\centering
\renewcommand{\arraystretch}{1.5} 
\caption{Spectral fitting parameters for the leptonic and lepto-hadronic models describing the multi-wavelength emission from the Swift~J1834–0846 region in the SNR scenario, for the single-source TeV and for the extended-component cases, and in the MWN scenario, for the point-like-component case. The columns list the model type, electron-to-proton ratio ($K_{\rm ep}$), electron and proton spectral indices ($\Gamma_{\rm e}$, $\Gamma_{\rm p}$), the break energy ($E_{\rm b}$), cutoff energies for electrons and protons ($E_{\rm e,cut}$, $E_{\rm p,cut}$), total energy in relativistic electrons ($W_{\rm e}$) and protons ($W_{\rm p}$), magnetic field strength ($B$), and Bayesian Information Criterion (BIC) values for statistical model comparison. Lower BIC values indicate more statistically favored models.}
\label{tab:Spec_model}
\vspace{0.3cm}

\resizebox{\textwidth}{!}{%
\begin{tabular}{lccccccccccc}
\hline
\hline
MODEL & \multicolumn{1}{l}{$K_{\rm ep}$} & $\Gamma_{e,1}$ & $\Gamma_{e,2}$ & $\Gamma_p$ & \begin{tabular}[c]{@{}c@{}}$E_{\mathrm{b}}$ \\ (TeV)\end{tabular} & \begin{tabular}[c]{@{}c@{}}$E_{e, \mathrm{cut}}$ \\ (TeV)\end{tabular} & \begin{tabular}[c]{@{}c@{}}$E_{p, \mathrm{cut}}$\\ (TeV)\end{tabular} & \begin{tabular}[c]{@{}c@{}}$W_e$\\ ($10^{50}$ erg)\end{tabular} & \begin{tabular}[c]{@{}c@{}}$W_p$\\ ($10^{50}$ erg)\end{tabular} & \begin{tabular}[c]{@{}c@{}}$B$\\ ($\mu$G)\end{tabular} & BIC \\ 
\hline
\multicolumn{12}{c}{Swift~J1834–0846 Region: Single-Source TeV Scenario - SNR} \\ 
\hline
Leptonic & --- & $2.83^{+0.10}_{-0.33}$ & $2.84^{+0.36}_{-0.09}$ & --- & $0.41^{+4.68}_{-0.39}$ & $20.93^{+9.89}_{-8.72}$ & --- & 2.26 & --- & $2.27^{+0.46}_{-0.37}$ & 36.82 \\
Lepto-hadronic & $10^{-1}$ & $1.87^{+0.16}_{-0.15}$ & $2.63^{+0.26}_{-0.29}$ & $2.19^{+0.14}_{-0.05}$ & $0.10^{+0.50}_{-0.07}$ & $3.21^{+5.79}_{-2.22}$ & $50.2^{+67.8}_{-26.9}$ & 0.034 & 1.96 & $38.4^{+34.4}_{-17.2}$ & 44.49 \\ 
Lepto-hadronic & $10^{-2}$ & $2.19^{+0.16}_{-0.14}$ & $2.28^{+0.52}_{-0.77}$ & $2.18^{+0.06}_{-0.04}$ & $2.46^{+11.48}_{-2.36}$ & $2.59^{+5.51}_{-1.57}$ & $49.5^{+27.7}_{-23.2}$ & 0.018 & 2.06 & $42.9^{+37.7}_{-19.9}$ & 44.05 \\
Lepto-hadronic & $10^{-3}$ & $2.45^{+0.14}_{-0.11}$ & $2.47^{+0.60}_{-0.85}$ & $2.16^{+0.03}_{-0.03}$ & $5.18^{+18.10}_{-5.08}$ & $2.88^{+9.82}_{-1.29}$ & $47.7^{+15.3}_{-12.2}$ & 0.0071 & 2.17 & $56.5^{+32.4}_{-23.9}$ & 44.37 \\
\hline
\multicolumn{12}{c}{Swift~J1834–0846 Region: Extended Component - SNR} \\ 
\hline
Leptonic & --- & $2.54^{+0.30}_{-0.32}$ & $3.14^{+0.17}_{-0.10}$ & --- & $0.05^{+0.33}_{-0.03}$ & $31.08^{+34.70}_{-15.85}$ & --- & 1.49 & --- & $2.42^{+0.88}_{-0.55}$ & 35.41 \\
Lepto-hadronic & $10^{-1}$ & $2.02^{+0.12}_{-0.15}$ & $2.30^{+0.50}_{-0.70}$ & $2.36^{+0.03}_{-0.05}$ & $0.88^{+34.6}_{-0.82}$ & $1.42^{+0.51}_{-0.22}$ & $315.3^{+216.2}_{-246.4}$ & 0.044 & 2.02 & $29.1^{+22.4}_{-10.8}$ & 42.41 \\ 
Lepto-hadronic & $10^{-2}$ & $2.25^{+0.14}_{-0.14}$ & $2.34^{+0.46}_{-0.76}$ & $2.28^{+0.05}_{-0.05}$ & $2.60^{+13.6}_{-2.54}$ & $2.03^{+8.36}_{-0.73}$ & $26.7^{+29.1}_{-18.9}$ & 0.018 & 2.27 & $39.8^{+31.1}_{-17.1}$ & 44.77 \\
Lepto-hadronic & $10^{-3}$ & $2.43^{+0.14}_{-0.10}$ & $2.40^{+0.57}_{-0.78}$ & $2.24^{+0.05}_{-0.07}$ & $7.08^{+10.6}_{-6.98}$ & $3.78^{+32.57}_{-3.58}$ & $12.9^{+16.2}_{-6.9}$ & 0.0052 & 2.26 & $70.3^{+36.7}_{-28.9}$ & 46.99 \\
\hline
\multicolumn{12}{c}{Swift~J1834–0846 Region: Point-Like TeV Component - MWN} \\ 
\hline
Leptonic & --- & $1.77^{+0.49}_{-0.50}$ & $3.59^{+0.27}_{-0.28}$ & --- & $1.73^{+0.49}_{-0.50}$ & $679.5^{+336.6}_{-519.1}$ & --- & 0.0048 & --- & $6.85^{+1.62}_{-1.29}$ & 15.86 \\
\hline
\hline
\end{tabular}%
}
\vspace{0.4cm}
\end{table*}

\section{Results and Discussion} \label{sec:results}

\begin{figure*} 
    \centering
    \begin{subfigure}{0.49\textwidth}
        \includegraphics[width=\linewidth]{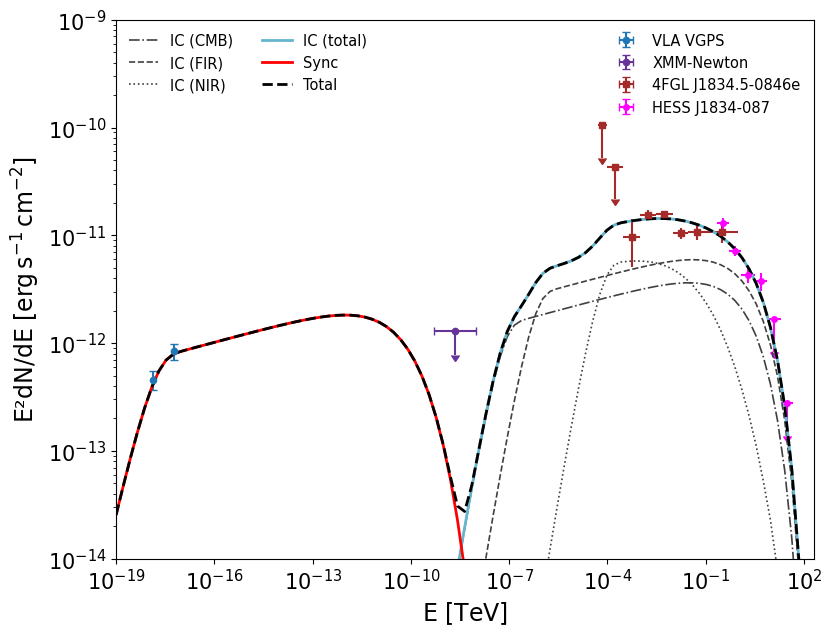}
        \caption{Leptonic model}
    \end{subfigure}
    \hfill
    \begin{subfigure}{0.49\textwidth}
        \includegraphics[width=\linewidth]{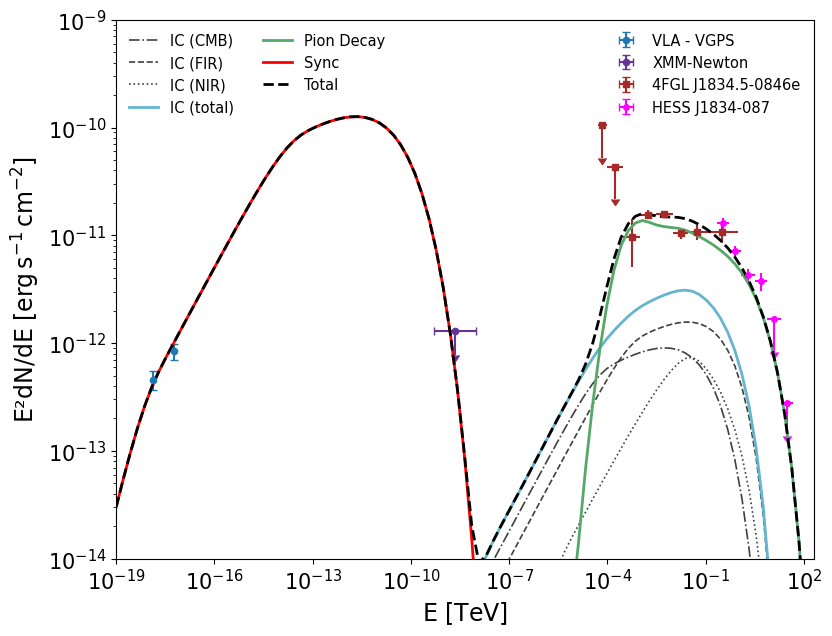}
        \caption{Lepto-hadronic model - $K_{\rm ep} = 10^{-1}$}
    \end{subfigure}
    \hfill
    \begin{subfigure}{0.49\textwidth}
        \includegraphics[width=\linewidth]{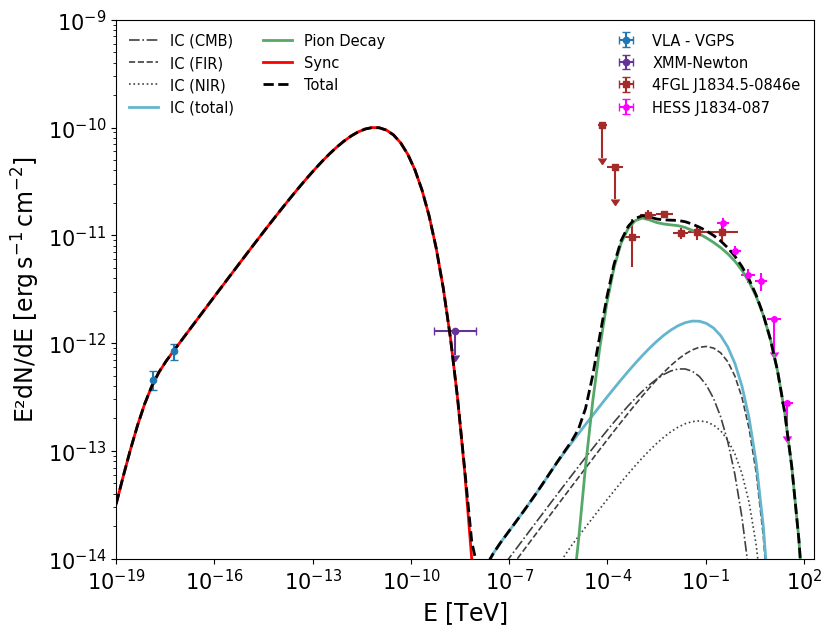}
        \caption{Lepto-hadronic model - $K_{\rm ep} = 10^{-2}$}
    \end{subfigure}
    \hfill
    \begin{subfigure}{0.49\textwidth}
        \includegraphics[width=\linewidth]{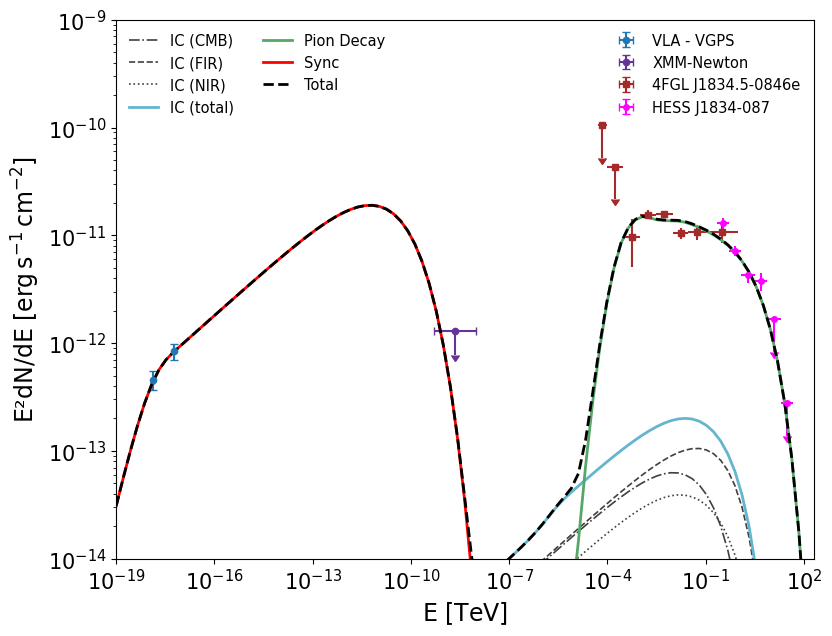}
        \caption{Lepto-hadronic model - $K_{\rm ep} = 10^{-3}$}
    \end{subfigure}
    \caption{Multi-band SED of the region of the magnetar Swift~J1834–0846 in the SNR scenario as a single source TeV, considering the total emission observed by \hess and \textit{Fermi}-LAT. The observational data include measurements in the radio \citep{1992AJ....103..943K, 2007ApJ...657L..25T}, X-ray \citep{2016ApJ...824..138Y}, GeV \citep{2017ApJS..232...18A, 2022ApJS..260...53A}, and TeV \citep{2018A&A...612A...1H} bands. The leptonic and lepto-hadronic models are combined with the three different electron-to-proton ratios. }
    \label{fig:Swift_PD_IC}
\end{figure*}

\subsection{Swift~J1834–0846 Region: Single-Source TeV Scenario}
\label{sec:modelling_Swift_single}

\vspace{0.1cm}

In this scenario, the TeV emission detected by \hess\ \citep{2018A&A...612A...1H} is treated as originating from a single source, and we examine whether the observed spectrum can be explained solely by emission from the SNR. Within the lepto-hadronic framework, we consider three representative values of the electron-to-proton ratio, $K_{\rm ep}  = 10^{-1}, 10^{-2}, 10^{-3}$, in order to assess their influence on the spectral fitting results.

The best-fit spectral parameters are summarized in Table~\ref{tab:Spec_model}, while the corresponding spectral energy distributions are presented in Figures~\ref{fig:Swift_PD_IC}-a (leptonic scenario) and Figures~\ref{fig:Swift_PD_IC}-b, -c and -d (lepto-hadronic scenario). These figures illustrate the relative contributions of the different radiation fields and emission mechanisms. In the purely leptonic model, gamma-ray emission around $\sim 10$~GeV is dominated by inverse Compton scattering of relativistic electrons off the NIR and FIR photon fields. Scattering off the FIR and CMB photon fields produces a broader spectral component and contributes to the gamma-ray flux up to energies of $\sim 10$~TeV.

In the lepto-hadronic models, the gamma-ray emission is predominantly driven by hadronic interactions. The contribution from IC scattering around $100$~GeV remains minor for $K_{\rm ep}  = 10^{-1}$ and $10^{-2}$, and becomes negligible in the case $K_{\rm ep}  = 10^{-3}$.The corresponding synchrotron emission is tightly constrained by the upper limit in the X-ray band, with its spectral peak located in the ultraviolet regime.

An important consideration is the total energy budget required by the particle populations to reproduce the observed emission (see Table~\ref{tab:Spec_model}). The total energy in relativistic electrons, $W_e$, ranges from $10^{48}$ to $10^{50}$~erg, while the total energy in relativistic protons, $W_p$, is around $\sim 10^{50}$~erg. These values are consistent with the commonly adopted assumption that roughly 10$\%$ of the kinetic energy released in a supernova explosion is converted into non-thermal particles, thereby supporting an SNR origin for the emission.

However, the magnetic field strength inferred in the purely leptonic model, $B \sim 2$~$\mu$G, argues against this scenario, as such a low value is atypical when compared to those inferred for other GeV–TeV–emitting SNRs \citep{2007Natur.449..576U, 2005ApJ...632..294B,2020pesr.book.....V}. For an SNR with an age in the range of $60-200$~kyr and interacting with surrounding H~I gas, the magnetic field is expected to be primarily shaped by environmental compression rather than by the strong amplification processes characteristic of young remnants. At an age of $\sim 60$ kyr, the remnant is still expected to exhibit a well-defined shell, with shocks strong enough to compress the interstellar magnetic field to values of $\sim 20-35$~$\mu$G \citep{2013ApJ...774...36C, 2021MNRAS.500.5177L}. By contrast, at $\sim 200$~kyr the remnant approaches the end of its evolutionary lifetime, and the magnetic field gradually relaxes toward the ambient interstellar value. Nevertheless, in H~I-rich environments the background magnetic field can remain relatively enhanced, reaching $\sim 15~\mu$G due to turbulent pressure within the cloud \citep[see e.g.,][]{2012ARA&A..50...29C}. In this context, the magnetic field strengths inferred in the lepto-hadronic models are more consistent with an SNR age of approximately $60$~kyr, favoring this evolutionary stage for the system.

Furthermore, the best-fit parameters of the leptonic model yield a spectral index change ($\Delta\Gamma_{e} = \Gamma_{e,2} - \Gamma_{e,1}$) for the accelerated electrons that deviates significantly from unity. In the canonical context of a SNR, radiative cooling steepens the electron spectrum by exactly one unit above the break energy, establishing a $\Delta\Gamma_{e} = 1$ \citep{1962SvA.....6..317K, 1970ranp.book.....P}. Reconciling the inferred $\Delta\Gamma_{e} \neq 1$ requires invoking fine-tuned mechanisms, for which there is no independent observational evidence or physical justification. Therefore, this non-canonical spectral break provides a robust, independent argument against a purely leptonic origin for the emission.

The advanced age of SNR~W41 ($\sim 60$~kyr) poses a tension with the high maximum proton energies ($\sim$ tens of TeV) required by the hadronic fit. This can be resolved if an inhomogeneous medium allows efficient shock acceleration in low-density regions before particles interact with adjacent dense clumps \citep{2012ApJ...744...71I}. Alternatively, these protons may be a fossil population accelerated during younger, faster evolutionary stages, currently confined near the remnant due to a small diffusion coefficient \citep{2009MNRAS.396.1629G}. Either scenario reconciles W41's advanced age with the hadronic model.

The models were further evaluated using the Bayesian Information Criterion (BIC) in order to identify the statistically favored scenario. Given that the purely leptonic models are disfavored by the unrealistically low magnetic field strengths they imply, we restricted the BIC comparison to the lepto-hadronic cases. Among these, the model with $K_{\rm ep} = 10^{-2}$ yields the smallest BIC value, indicating a statistical preference relative to the cases $K_{\rm ep} = 10^{-1}$ and $K_{\rm ep} = 10^{-3}$. The corresponding differences, $\Delta{\rm BIC} \approx 0.44$ and $\Delta{\rm BIC} \approx 0.32$, translate into relative probabilities\footnote{The relative probability that model 1 is statistically preferred over model 2 can be expressed as $P = 1/(1+\exp{(-\Delta{\rm BIC}/2)})$, where the difference is defined as $\Delta{\rm BIC} = {\rm BIC}_2 - {\rm BIC}_1$ \cite[see e.g.,][]{2007MNRAS.377L..74L, 2017ApJ...834..153Z}} of $P \approx 0.55$ and $P \approx 0.54$, respectively. However, according to the interpretative scale\footnote{According to the strength-of-evidence scale introduced by \cite{2008ConPh..49...71T}, differences in BIC values are interpreted as follows: $\Delta{\rm BIC} < 2$ corresponds to inconclusive evidence; $2$–$5$ indicates weak evidence; $5$–$10$ signifies moderate evidence; and $\Delta{\rm BIC} > 10$ represents strong evidence in favor of the model with the lower BIC value. \citep[see also][]{kass1995bayes}.} proposed by \cite{2008ConPh..49...71T}, these $\Delta{\rm BIC}$ values fall within the regime of inconclusive evidence. Therefore, statistical analysis does not provide sufficient support to definitively claim which model is preferred.


We now assess the physical consistency of the hypothesis that the SNR acts as the primary non-thermal emitter in this scenario, adopting as a reference the lepto-hadronic solution with $K_{\rm ep} = 10^{-1}$. In this model, the electron energy distribution is characterized by spectral indices $\Gamma_{e,1} \sim 1.9$ and $\Gamma_{e,2} \sim 2.2$, a break energy of $E_{\rm b} \sim 0.1$~TeV, and a cutoff energy of $E_{\rm e,cut} \sim 3.2$~TeV. These parameters produce a gamma-ray spectral peak around $100$~GeV. For a magnetic field strength of $B \sim 38~\mu$G, electrons with energies near the break energy have a synchrotron cooling time of $t_{\rm sync} \sim 87$~kyr, which is compatible with the estimated age of W41. This supports the interpretation that higher-energy electrons have already experienced substantial synchrotron losses. 

The proton distribution is described by a spectral index $\Gamma_p \sim 2.2$ and a high-energy cutoff at $E_{\rm p,cut} \sim 50$~TeV. Such a cutoff implies that protons are accelerated up to several tens of TeV, as required to account for the highest-energy gamma rays observed. The combination of relatively hard spectral indices and high cutoff energies is consistent with efficient particle acceleration via diffusive shock acceleration operating at the SNR shock front.

It is instructive to place the inferred magnetic field strength in the broader context of SNR studies. A field of $B \sim 38~\mu$G is compatible with expectations for an evolved remnant, or with a scenario in which the ambient interstellar magnetic field has been compressed by the passage of a strong shock \citep{2005JPhG...31R..95H, 2020pesr.book.....V}. Although such a field would be considered relatively low for a young SNR that exhibits clear evidence of magnetic-field amplification from non-thermal X-ray filaments, it remains fully consistent with the observational properties of SNR~W41, for which no definitive detection of non-thermal X-ray emission has been reported.

We also estimate the total magnetic energy content $W_{B}$, assuming a volume filling factor of unity, which provides an upper limit. For $B \sim 38~\mu\mathrm{G}$ and an SNR radius of $\sim 33$~pc \cite[corresponding to $27'$ at $4.2$~kpc; see][]{2007ApJ...657L..25T, 2025JApA...46...14G}, we obtain $W_{B} \sim 2.1 \times 10^{50}$~erg. This value exceeds the total energy in accelerated particles by approximately a factor of $W_{B}/(W_{e}+W_{p}) \sim 1.3 $ (see Table~\ref{tab:Spec_model}). Such an energy partition suggests that the remnant has likely evolved beyond its most efficient phase of particle acceleration and is currently in the late Sedov–Taylor stage, possibly approaching the transition to the radiative phase \citep{2013ApJ...773..138Y, 2018ApJ...855...59U}. In this evolutionary stage, particles can escape and radiative losses can exceed magnetic-field dissipation.

In summary, when the TeV emission is interpreted as originating from a single source, the SNR scenario with a lepto-hadronic particle spectrum characterized by $K_{\rm ep} = 10^{-1}$ provides a physically consistent and satisfactory description of the broadband emission observed in the Swift~J1834–0846 region.

\vspace{0.1cm}

\subsection{Swift~J1834–0846 Region: Extended and Point-Like TeV Components} \label{sec:modelling_Swift_point}

\vspace{0.1cm}

\begin{figure*} 
    \centering
    \begin{subfigure}{0.49\textwidth}
        \includegraphics[width=\linewidth]{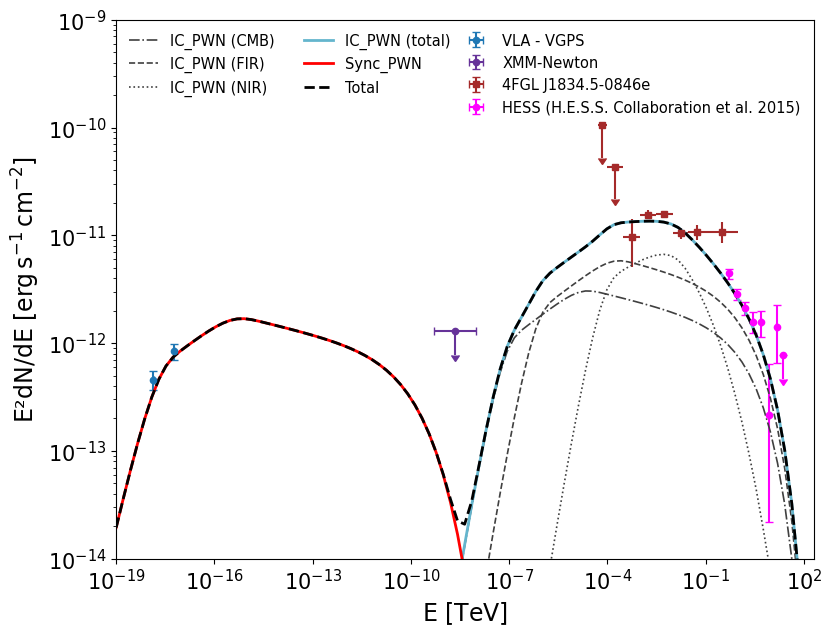}
        \caption{Leptonic model - Extended Component: SNR}
    \end{subfigure}
    \hfill
    \begin{subfigure}{0.49\textwidth}
        \includegraphics[width=\linewidth]{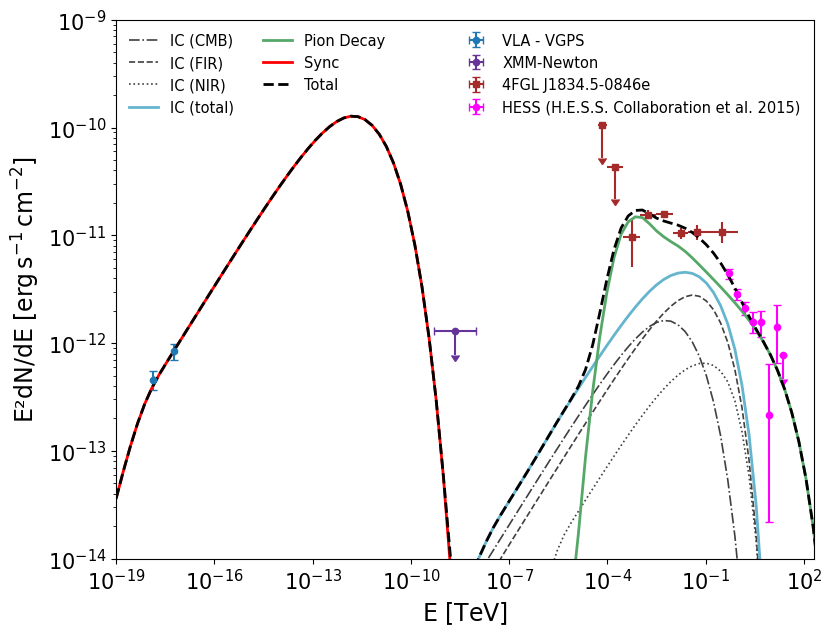}
        \caption{Lepto-hadronic model - $K_{\rm ep} = 10^{-1}$ - Extended Component: SNR}
    \end{subfigure}
    \hfill
    \begin{subfigure}{0.49\textwidth}
        \includegraphics[width=\linewidth]{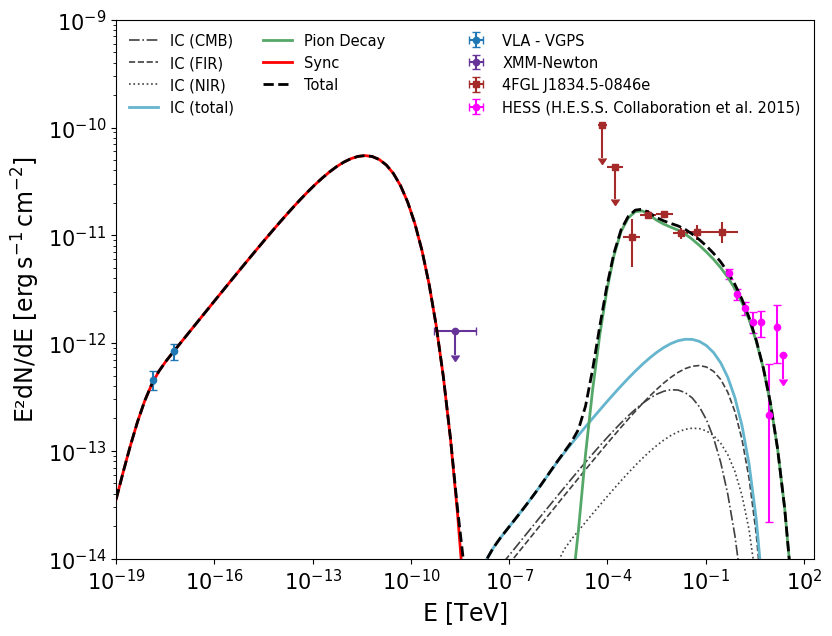}
        \caption{Lepto-hadronic model - $K_{\rm ep} = 10^{-2}$ - Extended Component: SNR}
    \end{subfigure}
    \hfill
    \begin{subfigure}{0.49\textwidth}
        \includegraphics[width=\linewidth]{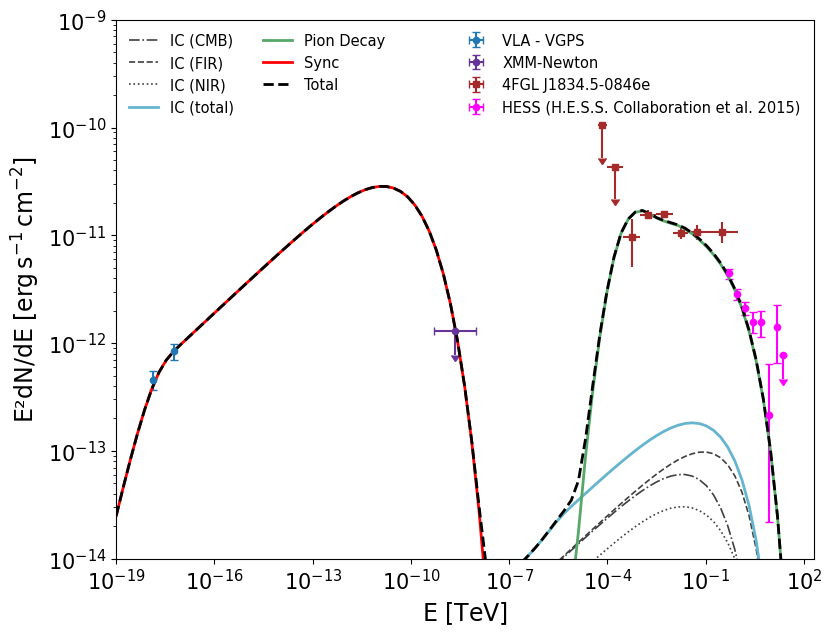}
        \caption{Lepto-hadronic model - $K_{\rm ep} = 10^{-3}$ - Extended Component: SNR}
    \end{subfigure}
    \hfill
    \begin{subfigure}{0.49\textwidth}
        \includegraphics[width=\linewidth]{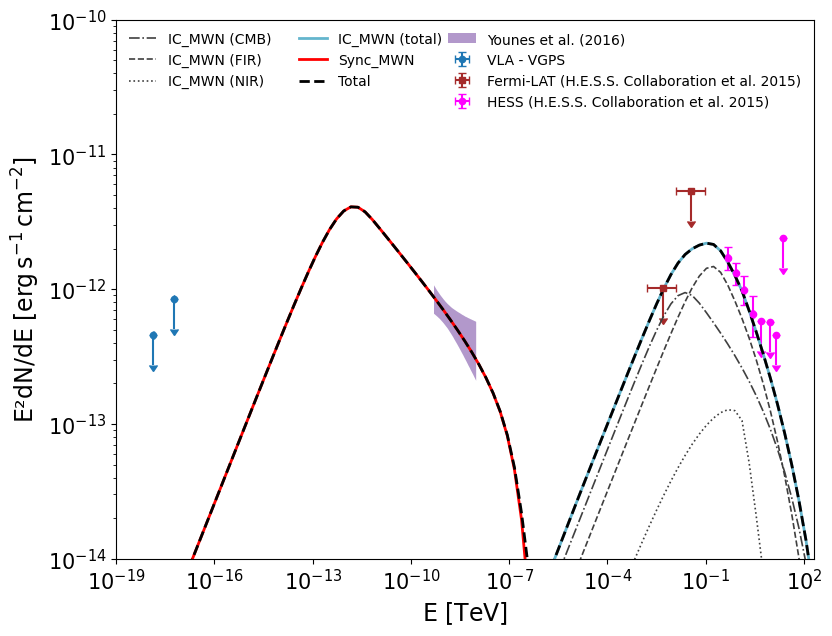}
        \caption{Leptonic model - Point-Like Component: MWN}
    \end{subfigure}
    \caption{Multi-band SED of the region around Swift~J1834–0846 assuming a two-component VHE morphology: an extended SNR component and a central point-like MWN component \citep{2015A&A...574A..27H}. The SNR is modeled with leptonic and lepto-hadronic particle spectra, while the MWN is modeled with an electron spectrum.}
    \label{fig:Swift_2comp}
\end{figure*}

In this section, we model the TeV emission as the superposition of two components: an extended component and a central point-like source. In this framework, the extended TeV emission is attributed to the SNR, while the point-like component is associated with the MWN. The TeV data were extracted from the normalized flux of the annular and central regions of HESS~J1834–087. Upper limits on the GeV emission from the point-like component were derived by introducing an additional point source on top of the extended GeV emission detected by \textit{Fermi}-LAT, adopting a confidence level of $3\sigma$ \citep{2015A&A...574A..27H}. The resulting fits are presented in Fig.~\ref{fig:Swift_2comp} and summarized in Table~\ref{tab:Spec_model}. For the SNR component, we considered both leptonic and lepto-hadronic models, whereas for the MWN component only leptonic models were explored. In the lepto-hadronic case, we adopted the same three values of the electron-to-proton ratio, $K_{\rm ep}$, as considered in the single-source scenario.

In the extended-SNR component scenario, the relative contributions of the various radiation fields and emission processes closely resemble those obtained in the single-source analysis. A notable difference is observed in the TeV band, where the extended component exhibits a softer spectral decline as a consequence of partitioning the TeV emission into two distinct components. Additionally, in the lepto-hadronic-$K_{\rm ep} = 10^{-1}$ model the relative contribution from the leptonic population is enhanced, although hadronic interactions continue to dominate the overall gamma-ray emission.

The inferred total non-thermal particle energies and magnetic field strengths in this two-component scenario are also comparable to those derived in the single-source case. Specifically, the total non-thermal energy requirements remain consistent with expectations for particle acceleration in a supernova explosion, and the magnetic field values inferred in the lepto-hadronic models are compatible with those expected for intermediate-age SNRs. In contrast, the magnetic fields required by the purely leptonic models are again very low and appear atypical for SNRs interacting with dense gas environments.

The ambient gas density interacting with the SNR may in fact be significantly higher than assumed in our baseline model, as the gamma-ray emission could originate from the interaction between SNR~W41 and the nearby giant molecular cloud (GMC) G23.0-0.4 \citep{2007ApJ...657L..25T, 2015ApJ...811..134S}. This GMC was first identified through CO observations by \cite{2006ApJ...643L..53A}, and the detection of OH maser emission at 1720 MHz by \cite{2013ApJ...773L..19F} provides compelling evidence for a physical interaction between the SNR shock and the molecular cloud. Moreover, the gamma-ray emission detected by \hess is spatially coincident with regions of high CO density \citep[see, e.g.,][]{2015A&A...574A..27H}, suggesting that the interaction between the SNR shock and the cloud may be responsible for the VHE emission. In this framework, hadrons accelerated at the SNR shock interact with dense target material, generating gamma rays via neutral pion decay, while the total energy in relativistic protons required to reproduce the GeV–TeV emission decreases as the ambient density increases.

This interpretation is further supported by the fact that numerous SNR–MC systems are known to be associated with extended GeV emission observed by \textit{Fermi}-LAT \citep{2009ApJ...706L...1A,2010ApJ...712..459A,2010ApJ...718..348A}. For W41, the estimated luminosity in the $0.1-100$~GeV range, $L_{\rm GeV} \sim 10^{35}$~erg~s$^{-1}$ (at 4.2 kpc), is consistent with other interacting SNRs detected by \textit{Fermi}-LAT \citep{2012APh....39...22T}. Within this framework, \cite{2015A&A...574A..27H} showed that the SNR–MC interaction model can reproduce the observed spectrum with a reasonable energy budget, while \cite{2012MNRAS.421..935L} proposed a diffusion model in which proton propagation as the remnant expands can account for both GeV and TeV emission, requiring a total injected energy of $W_p \sim 10^{50}$~erg. This value is in good agreement with the total energy estimates derived in our analysis.

In the point-like-MWN component scenario, the gamma-ray emission is dominated by inverse Compton scattering off the FIR and CMB photon fields, with only a minor contribution from the NIR component. The best-fit electron spectrum is characterized by spectral indices of $\Gamma_{\rm e,1} \approx 1.77$ and $\Gamma_{\rm e,2} \approx 3.59$. While the low-energy index $\Gamma_{\rm e,1}$ is consistent with values typically inferred for gamma-ray–emitting PWNe, the steeper high-energy index $\Gamma_{\rm e,2}$ reflects the requirement to reproduce the observed X-ray spectrum. The break energy, $E_{\rm e,b} \approx 2$ TeV, lies within the range commonly assumed for PWNe \citep{2014JHEAp...1...31T, 2024RAA....24g5016L}. These parameters yield a photon index of $\Gamma_{\gamma} \approx 2.29$ in the $100$~GeV–$1$~TeV energy range, consistent with the values generally reported for TeV-emitting PWNe \citep{2018A&A...612A...2H}.

The synchrotron component provides a good description of the lower-energy data, with the emission peaking in the infrared-to-optical band. The magnetic field strengths inferred from the synchrotron emission, $B \sim 6~\mu$G, are in good agreement with values derived for other PWNe \citep{2024RAA....24g5016L}.

Within this framework, the MWN is inferred to contain a total electron energy of $W_e \sim 5 \times 10^{47}$~erg. Adopting the magnetar’s characteristic age of $4.9$~kyr, its present-day spin-down luminosity of $\dot{E}_{\rm sd} \approx 2.1 \times 10^{34}$~erg~s$^{-1}$ \citep{2014ApJS..212....6O}, and a typical initial spin-down timescale for pulsars of $\tau_0 \sim 10^{3}$~yr \citep{2012MNRAS.427..415M, 2008ApJ...676.1210Z}, the total rotational energy released would be $E_{\rm sd} \sim 1.5 \times 10^{46}$ erg. This energy budget is clearly insufficient to account for the total energy stored in the relativistic electron population of the MWN.

However, if a much shorter initial spin-down timescale of $\tau_0 \sim 8-25$~yr is assumed, the total rotational energy released becomes sufficient to power the inferred electron population. In this case, the implied initial spin-down luminosity would be $\dot{E}_{\rm sd,0} \approx 8\times 10^{38}$~erg~s$^{-1}$, corresponding to an initial rotation period of $P_0 \lesssim 0.2$~s. Such a high luminosity is plausible in light of magnetar formation theories, which propose initial rotation periods on the order of milliseconds \citep{1993ApJ...408..194T, 2020MNRAS.494.4838L}.

We now assume that the nebula powered by Swift~J1834–0846 is physically associated with SNR~W41 and has an age of $\sim 60$ kyr, despite the magnetar’s much smaller characteristic age of $4.9$~kyr. Such discrepancies are known in other systems, e.g., PSR~B1757–24, where the true age exceeds the characteristic one due to non-constant braking parameters \citep{2000Natur.406..158G, 2013ApJ...768..144T}.

Using the observed extent of the diffuse X-ray emission \citep{2009ApJ...691.1707M} and adopting a distance of $4.2$~kpc, we infer a nebular radius of $\sim 3$~pc. For a magnetic field strength of $B \sim 6~\mu$G and a nebular size of this scale, the MWN is capable of accelerating electrons to very high energies. Moreover, the current spin-down luminosity, combined with an initial spin-down timescale of $\tau_0 \sim 10^{3}$~yr, is sufficient to supply the total electron energy required by our model. This implies a total rotational energy release of $E_{\rm sd} \sim 2.5 \times 10^{48}$~erg and an initial spin-down luminosity of at least $\dot{E}_{\rm sd,0} \sim 10^{38}$~erg~s$^{-1}$, corresponding to an initial rotation period of $P_0 \lesssim 0.3$~s.

We emphasize that, although the rotational energy losses alone can supply the overall energy budget of the MWN, additional particle outflows beyond the quiescent spin-down-powered wind may have contributed significantly to its evolution. These outflows are plausibly linked to the magnetar’s bursting activity, with a significant fraction of the injected energy originating from past outbursts. Such outflows are thought to be powered by the decay of both the dipole magnetic field and the much stronger internal magnetic field \citep[see e.g.,][]{2016RAA....16..143T, 2017MNRAS.464.4895G}.

\begin{figure*}
\centering
\includegraphics[width=0.6\textwidth]{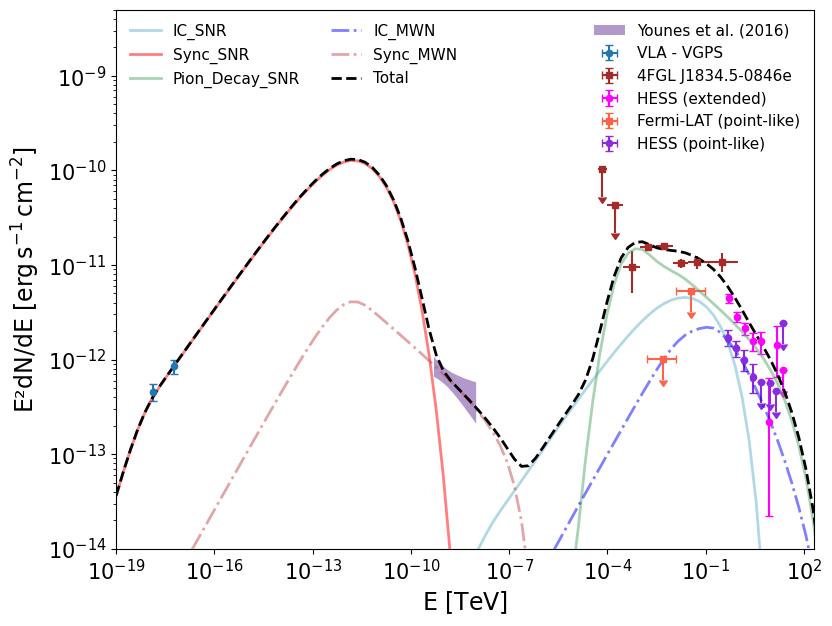}
\caption{Combined SEDs of the best-fit models for the region around Swift~J1834–0846, assuming a two-component VHE morphology consisting of an extended SNR component and a central point-like MWN component. The SNR emission is modeled using the lepto-hadronic-$K_{\rm ep} = 10^{-1}$ particle spectra, while the MWN emission is described by an {\it ECBPL} electron spectrum.}
\label{fig:swift_2comp}
\end{figure*}

Regarding the magnetic energy content of the MWN, assuming $B \sim 6~\mu$G and a radius of $3$~pc, the magnetic energy is $W_B \sim 5 \times 10^{45}$~erg, yielding a ratio $W_B/W_e \sim 0.09$ (see Table~\ref{tab:Spec_model}). This value ($\ll 1$) is consistent with what is typically reported for other PWNe, indicating a particle-dominated nebula \citep{1984ApJ...283..710K,2014JHEAp...1...31T}. At this evolutionary stage, around $60$~kyr, the PWN has likely already undergone its initial compression phase and is now experiencing a slow, but turbulent, re-expansion. The pressure and dynamics within the nebular region are therefore dominated by the relativistic particle population rather than by the magnetic field. A lower average magnetic field reduces synchrotron energy losses while extending the lifetime of high-energy electrons, thereby enhancing the relative contribution of IC emission. These conclusions, however, depend strongly on uncertain parameters such as distance and age.

Based on the BIC results for the extended component (see Table~\ref{tab:Spec_model}), we restrict the model comparison to the two lepto-hadronic scenarios, since the leptonic models are disfavored by the unusually low magnetic field strengths they require. Among the lepto-hadronic models, the case $K_{\rm ep} = 10^{-1}$ yields the lowest BIC value. The differences relative to the $K_{\rm ep} = 10^{-2}$ and $K_{\rm ep} = 10^{-3}$ configurations are $\Delta{\rm BIC} \approx 2.36$ and $4.58$, corresponding to relative probabilities of $P \approx 0.76$ and $0.91$, respectively. These results indicate that the $K_{\rm ep} = 10^{-1}$ model is moderately preferred over the alternative lepto-hadronic solutions, although the strength of evidence remains in the weak regime. 

For the point-like component, the leptonic scenario provides a physically consistent and satisfactory description of the observed broadband emission. The combined best-fit models for the two TeV components are presented in Fig.~\ref{fig:swift_2comp}.

\vspace{0.4cm}

\section{CTAO Contributions} \label{sec:CTAO}

\begin{figure*} 
    \centering
    \begin{subfigure}{0.49\textwidth}
        \includegraphics[width=\linewidth]{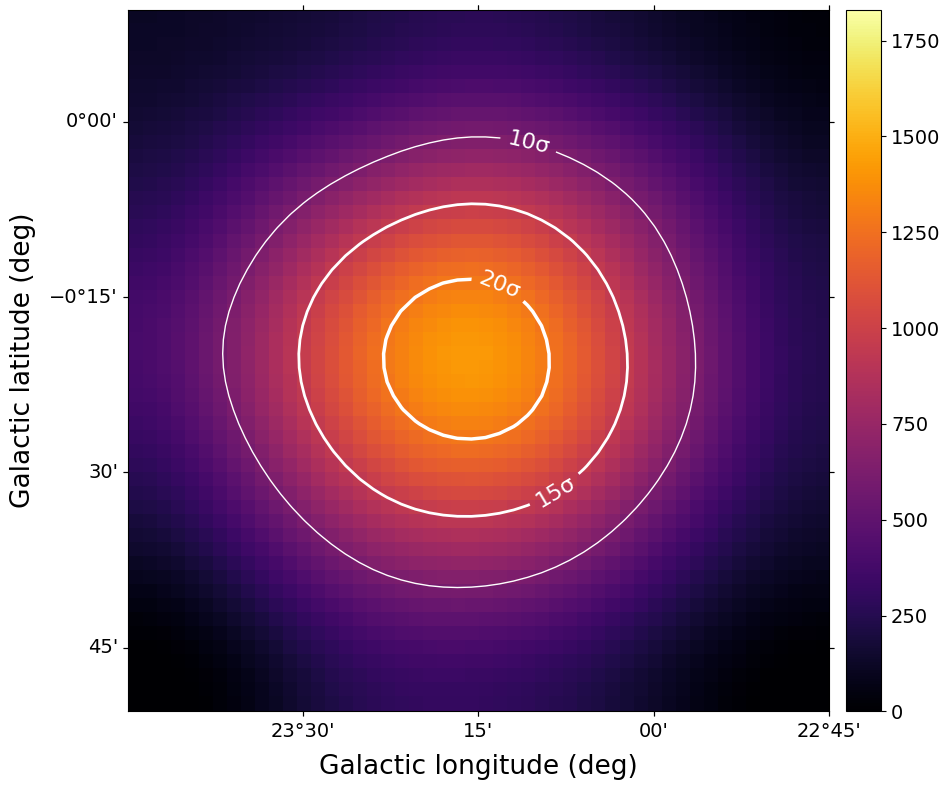}
    \end{subfigure}
    \hfill
    \begin{subfigure}{0.49\textwidth}
        \includegraphics[width=\linewidth]{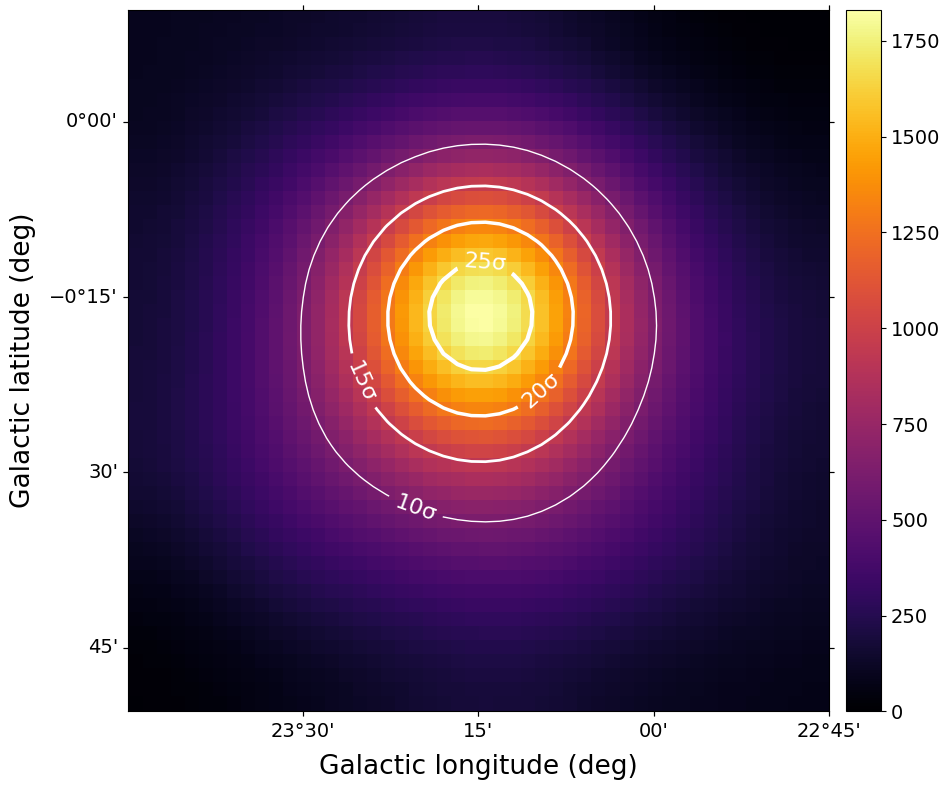}
    \end{subfigure}
    \hfill
    \begin{subfigure}{0.49\textwidth}
        \includegraphics[width=\linewidth]{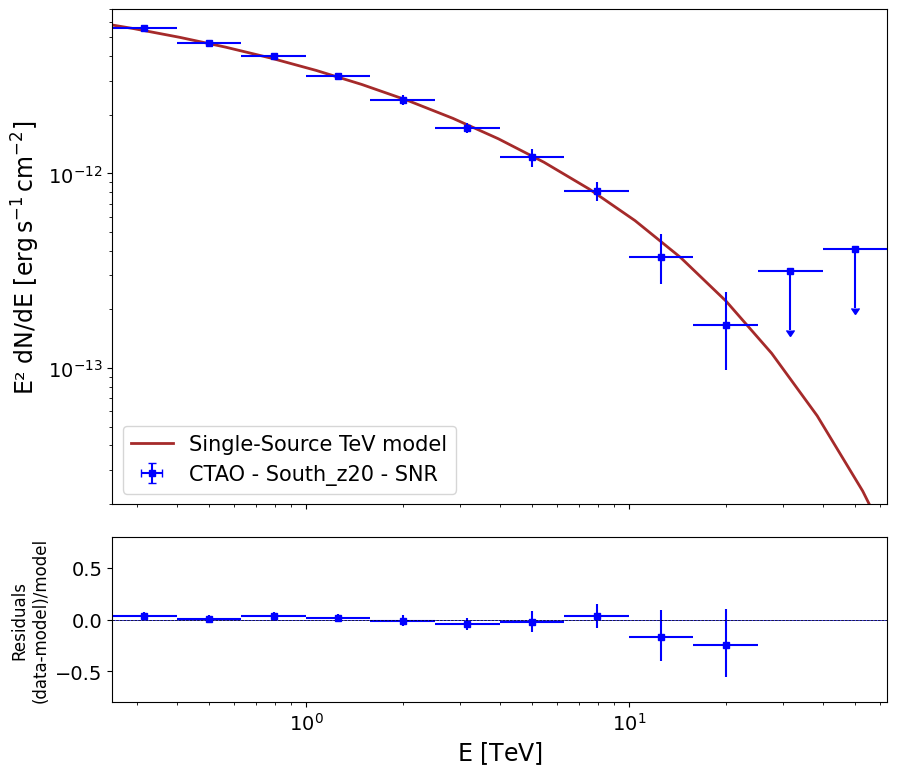}
        \caption{Single-Source TeV Model - SNR}
    \end{subfigure}
    \hfill
    \begin{subfigure}{0.49\textwidth}
        \includegraphics[width=\linewidth]{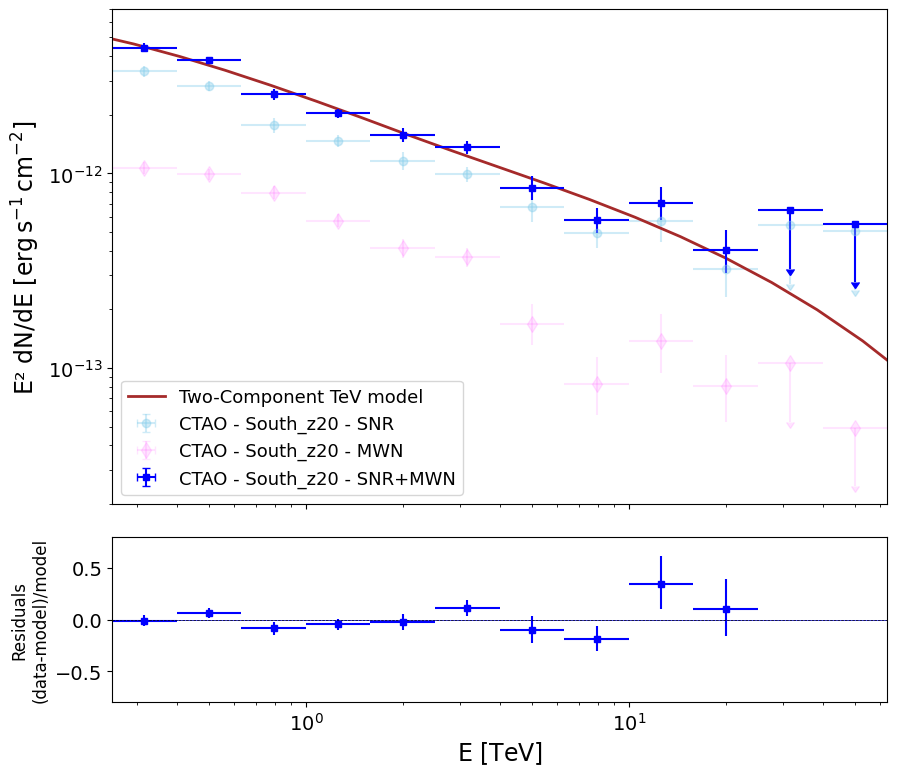}
        \caption{Two-Component TeV Model - SNR + MWN}
    \end{subfigure}
    \caption{Upper panels: Simulated CTAO maps of the VHE gamma-ray excess toward the Swift~J1834–0846 / SNR W41 region for the Single-Source TeV and Two-Component TeV scenarios. The images were generated using a spatial grid of $0.02^{\circ}$ and subsequently smoothed with a Gaussian kernel of $0.1^{\circ}$. Significance contours at 10, 15, 20, and 25$\sigma$ are overlaid as white solid lines. Lower panels: The corresponding simulated CTAO flux points together with the total SED for the Single-Source TeV and Two-Component TeV scenarios. For the Two-Component TeV configuration, the individual flux contributions from the SNR and MWN components are also shown separately, illustrating their respective roles in shaping the overall gamma-ray emission. Both the maps and the flux points were derived using the IRFs of the full southern array, optimized for a zenith angle of $z = 20^{\circ}$ and an exposure time of $t_{\rm obs} = 30$~hours.}
    \label{fig:ctao_simul}
\end{figure*}

\begin{table}
\centering
\renewcommand{\arraystretch}{1.6} 
\caption{Spatial models adopted for the 3D simulation of CTAO observations. These spatial models were taken from the morphological analysis of HESS~J1834--087 performed by H.E.S.S., as these measurements were obtained in an energy range consistent with that considered in our CTAO simulations \citep[see][]{2015A&A...574A..27H, 2018A&A...612A...1H}. The columns list the model type, Galactic longitude ($l$), Galactic latitude ($b$), and the source extension defined as the $1 \sigma$ Gaussian width ($\sigma_s$).}
\label{tab:Spatial_model}

\resizebox{0.48\textwidth}{!}{%
\begin{tabular}{lccc}
\hline
\hline
\begin{tabular}[c]{@{}l@{}}SPATIAL\\ MODEL\end{tabular} & $l$ ($^\circ$) & $b$ ($^\circ$) & $\sigma_s$ ($^\circ$) \\ \hline
\multicolumn{4}{c}{Single-Source TeV Scenario} \\ \hline
Gaussian - SNR & $23.263 \pm 0.016$ & $-0.332 \pm 0.017$ & $0.210 \pm 0.037$ \\ \hline
\multicolumn{4}{c}{Two-Component TeV Scenario} \\ \hline
Gaussian - SNR & $23.265 \pm 0.018$ & $-0.341 \pm 0.020$ & $0.223 \pm 0.017$ \\
Point-like - MWN & $23.243 \pm 0.007$ & $-0.265 \pm 0.008$ & --- \\ 
\hline
\hline
\end{tabular}%
}
\vspace{0.3cm}
\end{table}

The Cherenkov Telescope Array Observatory \citep[CTAO;][]{2019scta.book.....C} represents the next generation of ground-based instruments dedicated to VHE gamma-ray astronomy. It is specifically designed to investigate non-thermal processes in a wide range of astrophysical environments, including SNR, pulsar wind nebulae, active galactic nuclei, and other cosmic particle accelerators. As an open-access observatory in the VHE domain, CTAO will operate from two sites, located in the Northern and Southern hemispheres, thereby enabling full-sky coverage. The array will consist of telescopes of different sizes optimized to cover an energy range from a few tens of GeV up to beyond 100 TeV. With its improved angular and energy resolution, wide field of view, and flexible observing modes, CTAO is expected to deliver an order-of-magnitude enhancement in sensitivity compared to current-generation instruments, significantly advancing our ability to resolve source morphology, constrain spectral features, and probe the mechanisms of cosmic particle acceleration \citep{2019scta.book.....C}.

To evaluate the prospects of future observations with the CTAO in the Swift~J1834-0846 region, we performed dedicated 3D map simulations and derived the corresponding flux points using the \texttt{Gammapy} analysis framework \citep{2023A&A...678A.157D, acero_2025_17814297}. We adopted the Instrument Response Functions (IRFs) for the CTAO alpha configuration, which corresponds to the initial deployment phase of the array \citep{cherenkov_telescope_array_observatory_2021_5499840}. In particular, we used the IRFs of the full southern array, optimized for a zenith angle of $z = 20^\circ$. Simulated exposure time of $t_{\rm obs} = 30$~h was considered over the energy range $100$~GeV--$100$~TeV, consistent with the nominal sensitivity domain of the CTAO southern site. This configuration was adopted as it provides near-optimal performance for the source’s sky position, ensuring favorable zenith angles and maximizing the sensitivity of the southern array.

The emission from the region was modeled under the two scenarios investigated in this work: the Single-Source TeV scenario (SNR) and the Two-Component TeV scenario (SNR+MWN). In the Single-Source TeV case, we adopted as the spectral model the combined IC and neutral-pion decay emission arising from a lepto-hadronic particle distribution with an electron-to-proton ratio of $K_{\rm ep} = 10^{-1}$. For the Two-Component TeV scenario, the region was modeled by including both contributions: the SNR, described by the lepto-hadronic-$K_{\rm ep} = 10^{-1}$ particle spectrum, and the MWN, characterized by a leptonic model (see Table~\ref{tab:Spec_model}). These model selections are motivated by their statistical preference in the spectral fits, as well as by their physical consistency within the proposed interpretation. Furthermore, we assumed a Gaussian spatial template to model the SNR emission in both scenarios, while the MWN component was described using a point-like spatial model (see Table~\ref{tab:Spatial_model}). The parameters of these spatial distributions are adopted from the morphological analyses reported for HESS~J1834--087 under the assumptions of single- and two-component configurations \citep[see][]{2015A&A...574A..27H, 2018A&A...612A...1H}.

Based on the simulated observations, we find that CTAO will detect the Swift~J1834–0846/W41 region with a statistical significance of $\sim 47\sigma$ in the single-source scenario and  $\sim 39\sigma$ in the two-component scenario. These results demonstrate that the gamma-ray emission from the region can be robustly detected with an exposure time of $t_{\rm obs} = 30$~hours.

Figure~\ref{fig:ctao_simul} shows the simulated VHE gamma-ray excess maps, together with the corresponding flux points and the assumed best-fit spectral models for each scenario. The maps indicate that CTAO will be capable of discriminating between the two configurations. In the single-source case, the emission attributed to the SNR appears more spatially extended and exhibits a lower peak gamma-ray excess. In contrast, in the two-component scenario, the presence of the MWN, modeled as a point-like source, introduces a centrally concentrated excess, leading to higher significance levels toward the center of the emission region. This behavior is also reflected in the overlaid significance contours (solid white lines), which show a more pronounced central enhancement in the two-component configuration.

The reconstructed flux points closely reproduce the input spectral model, with relative residuals ([\text{data} - \text{model}]/\text{model}) below $0.5$ across the explored energy range. This demonstrates that the CTAO can reliably recover the intrinsic spectral shape of the source. Notably, even in the presence of spatial overlap between components, CTAO retains sufficient angular resolution and sensitivity to identify the point-like MWN contribution superimposed on the extended SNR emission (see Fig~\ref{fig:ctao_simul}-b).

In comparison with previous measurements, the CTAO energy flux uncertainties are substantially reduced, particularly at lower energies, highlighting the significant gain in sensitivity and spectral precision relative to current facilities such as \hess\@. In addition, the reconstructed spectrum extends beyond $\sim 10$~TeV, enabling tighter constraints on the high-energy cutoff of the proton distribution. This improvement is especially relevant for longer exposure times, which further enhance the capability of CTAO to probe the maximum particle energies achieved in the system.

In summary, observations with the CTAO in this region will not only provide a high-significance detection, but also sufficient sensitivity to discriminate between the alternative TeV-component interpretations on morphological criteria. Moreover, it will provide significantly improved constraints on the maximum energies attained by the particles responsible for the observed gamma-ray spectra.

\section{Summary and conclusions}    \label{conc}

We have investigated the non-thermal emission in the vicinity of 
Swift~J1834--0846/W41 by modeling its multiwavelength spectral energy 
distribution within leptonic and lepto-hadronic frameworks. Through MCMC 
sampling implemented in \texttt{Naima}, we constrained the energy distributions of the underlying particle populations responsible for the observed radio-to-TeV radiation, isolating the respective contributions of synchrotron emission, IC scattering, and neutral-pion decay.

Motivated by morphological studies of HESS~J1834--087 indicating a potential 
two-component structure, we explored two alternative scenarios. In the first, the entire TeV signal is attributed to a single extended source in order to assess whether the SNR alone can reproduce the broadband spectrum. Both purely leptonic and lepto-hadronic SNR models were examined within this context. While the leptonic model can reproduce the overall SED, it requires an exceptionally low magnetic field of $B \sim 2~\mu$G, significantly below 
typical values inferred for GeV--TeV SNRs, thereby disfavoring this 
interpretation on physical grounds. In contrast, the lepto-hadronic solutions yield magnetic field strengths of $B \sim 40~\mu$G, consistent with expectations for an evolved remnant in which the interstellar magnetic field has been compressed by the passage of the SNR shock.

For the preferred lepto-hadronic configurations, the inferred non-thermal 
energy budgets of $W_{\rm e} \sim 10^{48}$--$10^{49}$~erg and 
$W_{\rm p} \sim 10^{50}$~erg are consistent with the canonical picture in 
which approximately 10\% of the supernova kinetic energy is converted into 
non-thermal particles. The estimated magnetic energy content slightly exceeds the total energy stored in accelerated particles, suggesting that W41 is likely in a late Sedov--Taylor evolutionary stage, possibly approaching the transition to the radiative phase. The proton spectrum is characterized by a spectral index of $\Gamma_{\rm p} \sim 2.2$ and a high-energy cutoff at $E_{\rm p,cut} \sim 50$~TeV, consistent with efficient diffusive shock acceleration operating at the SNR shock front. Among the lepto-hadronic models, the configuration with $K_{\rm ep} = 10^{-2}$ yields the lowest BIC value, although the differences relative to the $K_{\rm ep} = 10^{-1}$ and $K_{\rm ep} = 10^{-3}$ cases fall within the regime of inconclusive evidence. Overall, when the TeV emission is treated as originating from a single source, the SNR with a lepto-hadronic particle distribution characterized by $K_{\rm ep} = 10^{-1}$ provides a physically consistent and satisfactory description of the data.

In the second scenario, the TeV emission is decomposed into an extended 
component associated with the SNR and a central point-like component attributed to the MWN powered by Swift~J1834--0846. For the extended SNR component, both leptonic and lepto-hadronic models were again tested, yielding results that closely resemble those obtained in the single-source case. Hadronic interactions dominate the gamma-ray output in the lepto-hadronic solutions, while purely leptonic models again require unrealistically low magnetic fields. The total non-thermal energy requirements remain within the range expected for supernova 
remnants and are compatible with a scenario in which SNR~W41 interacts with the nearby giant molecular cloud G23.0--0.4, thereby enhancing the target density for proton--proton collisions. This interpretation is further supported by the spatial coincidence between the H.E.S.S. gamma-ray emission and regions of high CO density, as well as by the detection of OH maser emission at 1720~MHz indicative of an SNR--molecular cloud interaction. Among the lepto-hadronic models for the extended component, the $K_{\rm ep} = 10^{-1}$ configuration is moderately preferred over the alternatives, with relative probabilities of $P \approx 0.76$ and $P \approx 0.91$ with respect to the $K_{\rm ep} = 10^{-2}$ and $K_{\rm ep} = 10^{-3}$ cases, respectively.

For the central point-like component, a leptonic model provides a coherent 
description of the broadband emission. The derived electron spectrum is 
characterized by spectral indices of $\Gamma_{\rm e,1} \approx 1.77$ and 
$\Gamma_{\rm e,2} \approx 3.59$, with a break energy of $E_{\rm b} \approx 2$~TeV and a magnetic field strength of $B \sim 6~\mu$G, consistent with values typically reported for TeV-emitting pulsar wind nebulae. The steeper high-energy spectral index indicates the need to reproduce the observed X-ray spectrum. The total electron energy implied by this model, $W_{\rm e} \sim 5 \times 10^{47}$~erg, together with the inferred nebular radius of approximately 3~pc, suggests a short initial spin period for the magnetar of $P_0 \lesssim 0.2$~s, which remains compatible with current theoretical expectations for magnetar formation. Furthermore, the ratio $W_{\rm B}/W_{\rm e} \sim 0.09$ confirms that the MWN is a particle-dominated nebula, consistent with what is typically reported for evolved pulsar wind nebulae. Within this two-component framework, the extended TeV emission is therefore primarily hadronic in origin, while the central excess is naturally explained by leptonic radiation from the MWN.

We also evaluated the observational prospects with the CTAO. Our simulations 
indicate that CTAO will detect the Swift~J1834--0846/W41 region with statistical significances of approximately $47\sigma$ and $39\sigma$ in the single-source and two-component scenarios, respectively, for an exposure time of 30~h. More importantly, the improved angular resolution will enable a clearer morphological discrimination between the two configurations, as the centrally concentrated excess associated with the MWN becomes distinguishable from the more extended SNR emission. The significance contours obtained from the simulated maps already illustrate this capability, showing a more pronounced central enhancement in the 
two-component scenario. Spectrally, CTAO will substantially reduce flux 
uncertainties and extend the measurable spectrum beyond $\sim$10~TeV, thereby placing significantly tighter constraints on the maximum energies of accelerated electrons and protons and enabling a more robust characterization of the high-energy cutoff of the proton distribution.

In conclusion, our multiwavelength study supports a complex non-thermal 
environment in the Swift~J1834--0846/W41 region. Our analysis indicates that a lepto-hadronic SNR model with $K_{\rm ep} = 10^{-1}$ provides the most 
physically plausible description of the extended TeV emission, while a leptonic MWN component naturally accounts for the central point-like source. Although current data do not uniquely favor one scenario, future CTAO observations will be decisive in clarifying the TeV morphology and in establishing the dominant particle-acceleration mechanisms operating in this system.

\section*{Acknowledgements}
We thank the anonymous referee for their valuable suggestions and comments. M.F.S. and V.d.S. thank Fundação de Amparo à Pesquisa do Estado de São Paulo (FAPESP, grants No. 2025/05794-2 and No. 2021/01089-1) for the financial support. V.d.S. is supported by CNPq through grant number 308837/2023-1. R.C.A. acknowledge the financial support of the NAPI “Fenômenos Extremos do Universo” of Fundação de Apoio à Ciência, Tecnologia e Inovação do Paraná. R.C.A. research is supported by CAPES/Alexander von Humboldt Program (grant No. 88881.800216/2022-01), CNPq (grant Nos. 310448/2021-2 and 4000045/2023-0), Araucária Foundation (grant Nos. 698/2022 and 721/2022) and FAPESP (grant No. 2021/01089-1). R.C.A. also acknowledges the support of L’Oreal Brazil, with the partnership of ABC and UNESCO in Brazil. The research used Gammapy, a Python package developed by the community for TeV gamma-ray astronomy \citep{2017ICRC...35..766D, 2023A&A...678A.157D}, accessible at \href{https://www.gammapy.org}{https://www.gammapy.org}. In addition, we used the instrument response functions for the Cherenkov Telescope Array Observatory (CTAO) provided by the CTA Consortium and CTAO. For detailed information on these instrument response functions, see \href{https://www.ctao-observatory.org/science/cta-performance}{https://www.ctao-observatory.org/science/cta-performance} \citep[version prod5 v0.1;][]{cherenkov_telescope_array_observatory_2021_5499840}.

\appendix



\bibliographystyle{elsarticle-harv} 
\bibliography{references}






\end{document}